\documentclass{mem}
\usepackage{natbib}\usepackage{txfonts}\usepackage{balance}
\usepackage{graphicx}
\usepackage[a4paper,breaklinks,dvipdfm]{hyperref}
\idline{75}{282}
\begin{document}

\def\la{\;
\raise0.3ex\hbox{$<$\kern-0.75em\raise-1.1ex\hbox{$\sim$}}\; }
\def\ga{\;
\raise0.3ex\hbox{$>$\kern-0.75em\raise-1.1ex\hbox{$\sim$}}\; }

\title{
White dwarf constraints on a varying $G$
}

\author{
E. \,Garc\'\i a--Berro\inst{1,2} 
\and 
S. \, Torres\inst{1,2}
\and
L. G. \,Althaus\inst{3}
\and
A. H. \,C\'orsico\inst{3}
\and 
\\P. \, Lor\'en--Aguilar\inst{4}
\and
A. D. \, Romero\inst{5}
\and 
J. Isern\inst{6,2}
}

\offprints{E. Garc\'\i a--Berro}

\institute{Departament de F\'\i sica Aplicada, 
           Universitat Polit\`ecnica de Catalunya,
           c/Esteve Terrades, 5,  
           08860 Castelldefels, 
           Spain
\and
           Institute for Space Studies of Catalonia,
           c/Gran Capit\`a 2--4, Edif. Nexus 104,   
           08034  Barcelona, 
           Spain
\and
           Facultad de Ciencias Astron\'omicas y Geof\'{\i}sicas,  
           Universidad  Nacional de La Plata,
           Paseo del  Bosque s/n,  
           (1900) La Plata, 
           Argentina
\and
           School of Physics, 
           University of Exeter, 
           Stocker Road, Exeter EX4 4QL,
           United Kingdom
\and
           Departamento de Astronomia,
           Universidade Federal do Rio Grande do Sul,\\
           Av. Bento Goncalves 9500, Porto Alegre 91501-970, RS, 
           Brazil
\and
           Institut de Ci\`encies de l'Espai, CSIC, 
           Campus UAB, Facultat de Ci\`encies, Torre C-5, 
           08193 Bellaterra, 
           Spain\\
\email{enrique.garcia-berro@upc.edu}
}

\authorrunning{Garc\'\i a--Berro et al.}
\titlerunning{White dwarf constraints on a varying $G$}

\abstract{A  secular  variation  of  $G$ modifies  the  structure  and
  evolutionary time scales of white dwarfs.  Using an state-of-the-art
  stellar  evolutionary code,  an up-to-date  pulsational code,  and a
  detailed population  synthesis code we demonstrate  that the effects
  of a  running $G$ are obvious  both in the properties  of individual
  white  dwarfs,  and in  those  of  the  white dwarf  populations  in
  clusters.  Specifically,  we show that the  white dwarf evolutionary
  sequences depend on  both the value of $\dot G/G$,  and on the value
  of $G$  when the  white dwarf was  born.  We show  as well  that the
  pulsational  properties of  variable  white dwarfs  can  be used  to
  constrain  $\dot G/G$.   Finally,  we also  show  that the  ensemble
  properties of  of white dwarfs in  clusters can also be  used to set
  upper bounds to $\dot G/G$.  Precisely, the tightest bound --- $\dot
  G/G  \sim  -1.8 10^{-12}$~yr$^{-1}$  ---  is  obtained studying  the
  population  of the  old,  metal-rich, well  populated, open  cluster
  NGC~6791. Less stringent upper limits  can be obtained comparing the
  theoretical results  obtained taking into  account the effects  of a
  running $G$ with the measured rates  of change of the periods of two
  well  studied pulsating  white dwarfs,  G117--B15A and  R548.  Using
  these   white   dwarfs   we    obtain   $\dot   G/G\sim   -1.8\times
  10^{-10}$~yr$^{-1}$,      and      $\dot     G/G\sim      -1.3\times
  10^{-10}$~yr$^{-1}$, respectively,  which although  less restrictive
  than  the previous  bound, can  be  improved measuring  the rate  of
  change of the period of massive white dwarfs.
\keywords{Stars -- White dwarfs -- Gravity}}

\maketitle{}

\section{Introduction}

General Relativity  is the favorite  theory of gravitation, and  it is
based on the  equivalence principle, and in the end  on the assumption
that the  gravitational constant,  $G$, is indeed  constant.  However,
this is just a hypothesis which  needs to be verified.  In fact, there
are several  modern grand-unification  theories that predict  that the
value of  $G$ is  a varying  function of  a low-mass  dynamical scalar
field \citep{LAea,mio}.  Hence,  we expect that if  these theories are
true the  gravitational constant  should experience slow  changes over
cosmological  timescales. In  recent years,  several constraints  have
been placed on the variation of the fine structure constant, and other
interesting constants  of nature  --- see \cite{uzan},  and \cite{mio}
for extensive reviews.   However, very few works have  been devoted to
study a  hypothetical variation of $G$.  The most tight bounds  on the
variation  of $G$  are those  obtained using  Lunar Laser  Ranging ---
$\dot{G}/G  =  (0.2\pm0.7)\times 10^{-12}$~yr$^{-1}$  \citep{H10}  ---
solar    asteroseismology   ---    $\dot    G/G   \simeq    -1.6\times
10^{-12}$~yr$^{-1}$ \citep{Demarque} ---  and Big Bang nucleosynthesis
--- $-0.3  \times  10^{-12}$~yr$^{-1}   \la  \dot{G}/G  \la  0.4\times
10^{-12}$~yr$^{-1}$ \citep{CO4,B05}.   Nevertheless, both  Lunar Laser
Ranging and asteroseismological bounds  are eminently local, while Big
Bang limits  are model-dependent.   At intermediate  cosmological ages
the Hubble  diagram of Type  Ia supernovae has  also been used  to put
constraints on  the rate  of change  of $G$,  but the  constraints are
somewhat  weaker $\dot  G/G\la 1\times  10^{-11}$~yr$^{-1}$ at  $z\sim
0.5$~\citep{SNIa,IJMPD}. In  this work we  summarize why on  how white
dwarfs can be used to place constraints  on the rate of variation of a
rolling $G$.

\section{White dwarf cooling times}

\begin{figure}[t!]
\resizebox{\hsize}{!}{\includegraphics[clip=true]{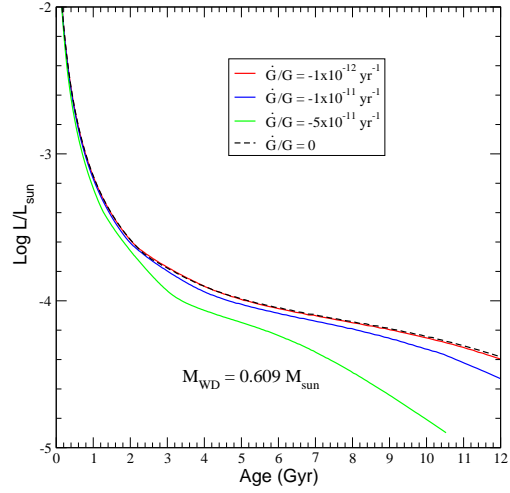}}
\caption{\footnotesize  Surface  luminosity  versus  age  for  several
  $0.609\, M_\odot$  white dwarf sequences, adopting  different values
  of $\dot{G}/G$.}
\label{fig1}
\end{figure} 

A  slowly rolling  $G$ affects  both the  cooling timescales  of white
dwarfs \citep{gold,gnew} and those of their progenitors \citep{Weiss}.
Consequently, the ages determined from the color-magnitude diagrams of
globular or open  clusters --- namely, the  main-sequence turn-off age
and the age determined from the termination of the white dwarf cooling
sequence ---  change accordingly, and  depend on the precise  value of
$\dot G/G$ \citep{JCAP}.  This allows to put upper bounds  on the rate
of variation of $G$.  To quantify the  effects of a varying $G$ on the
derived ages,  we computed  the main sequence  evolution of  two model
stars of  1.0 and $2.0\,  M_{\odot}$ considering three values  for the
rate    of    change    of   $G$,    namely    $\dot{G}/G=-5    \times
10^{-11}$~yr$^{-1}$,  $\dot{G}/G=-1  \times  10^{-11}$~yr$^{-1}$,  and
$\dot{G}/G=-1  \times   10^{-12}$~yr$^{-1}$.   All   the  evolutionary
calculations  were done  using the  {\tt LPCODE}  stellar evolutionary
code \citep{Althaus10,Renedo10},  appropriately modified to  take into
account  the effect  of a  varying $G$.   Despite the  small rates  of
change of  $G$ adopted here,  the evolution of white  dwarf progenitor
stars  is  severely  modified.   The evolutionary  timescales  can  be
modelled using rather  simple arguments.  In particular,  it turns out
that the  main sequence  lifetimes when  a varying  $G$ is  adopted is
\citep{JCAP}:

\begin{equation}
\tau_{\rm MS}=\frac{1}{\gamma\left|\frac{\dot G}{G}\right|}
\ln\left[\gamma\left|\frac{\dot G}{G}\right|\left(\frac{G_0}{G_i}\right)^\gamma
\tau_{\rm MS}^0+1\right].
\label{fit}
\end{equation}

\noindent with $\gamma=3.6$. The effect of  a varying $G$ on the white
dwarf cooling  times is displayed  in Fig.~\ref{fig1}. As can  be seen,
the cooling  timescales are  considerably modified, being  the cooling
accelerated  in the  case of  $\dot  G<0$ \citep{gnew}.   This can  be
explained  easily.   A   smaller  value  of  $G$   implies  a  smaller
gravitational force,  and thus a  smaller degeneracy (and  density) is
needed  to balance  gravity.  Hence,  for $\dot  G<0$ the  white dwarf
expands as  it evolves, and  the cooling is accelerated.   The cooling
track shown in Fig.~\ref{fig1} is a representative example of a set of
white dwarf  cooling sequences  which incorporate the  most up-to-date
physical  inputs.   Specifically,  these  cooling  sequences  consider
$^{22}$Ne    diffusion    and    its   associated    energy    release
\citep{nature,Althaus10,GB08},  and together  with  the main  sequence
lifetimes  given by  Eq.~(\ref{fit}) allow  to derive  an age  for any
cluster, and  for each  value of  $\dot G/G$.   Moreover, the  grid of
models has  been computed for several  initial values of $G$,  to take
into account that the evolutionary value of $G$ must match its present
value.   With these  sequences  the effect  of a  running  $G$ on  the
color-magnitude diagram of any cluster can be studied.

\begin{figure}[t!]
\resizebox{\hsize}{!}{\includegraphics[clip=true]{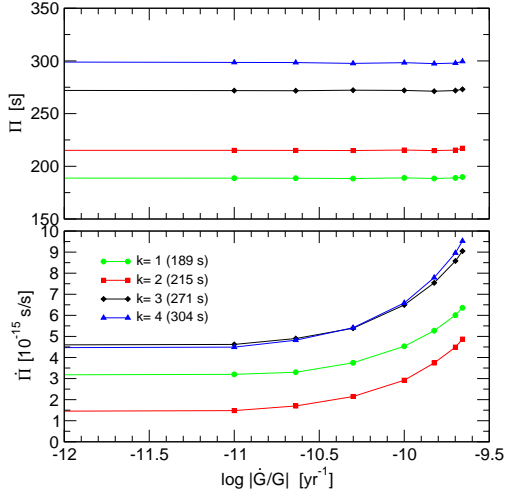}}
\caption{\footnotesize Upper  panel: periods  of the several  modes of
  G117$-$B15A as a function of the value of $|\dot G/G|$ with $\dot{G}
  <  0$. Lower  panel:  period  derivatives of  the  same  modes as  a
  function of the secular rate of change of $G$.}
\label{fig2}
\end{figure}

We chose  to employ the  old, metal-rich, well populated  open cluster
NGC~6791, for which the ages derived from main sequence stars and from
the termination of the degenerate sequence agree very well in the case
of a constant  $G$.  When a varying $G$ is  adopted, the resulting age
of NGC~6791  is modified, but then  the position of the  main sequence
turn-off  in   the  color-magnitude  diagram  is   also  significantly
different  if the  same  distance modulus  is  adopted.  However,  the
distance  modulus derived  using  an independent  and reliable  method
(eclipsing binaries)  which does  not make  use of  theoretical models
turns out to  be $13.46\pm 0.1$ \citep{Grundahl}.   Thus, large errors
in  the distance  modulus seem  to  be quite  implausible.  Given  the
uncertainty in the distance modulus  ($\simeq 0.1^{\rm mag}$), and the
measured  value,  an   upper  limit  to  $\dot  G/G$   can  be  placed
\citep{JCAP}.    Since   $\Delta    t_{\rm   MSTO}/\Delta   (m-M)_{\rm
F606W}\approx 4$~Gyr/mag,  the maximum age difference  with respect to
the case in  which a constant $G$ is adopted  is $\sim 0.4$~Gyr, which
translates   into   an   upper    bound   $\dot   G/G\sim   -1.8\times
10^{-12}$~yr$^{-1}$.  This upper limit considerably improves the other
existing  upper  bounds  to  the  rate of  variation  of  $G$  and  is
equivalent to the upper limit set by helioseismology.

\begin{figure}[t!]
\resizebox{\hsize}{!}{\includegraphics[clip=true]{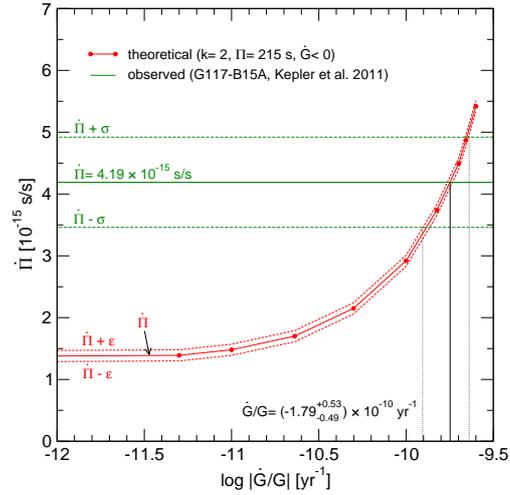}}
\caption{\footnotesize Rate of temporal  variation of the 215~s period
  of G117$-$B15A  as a function  of $\dot{G}/G$, red solid  line.  The
  observational  value  of  the  rate  of change  of  the  period  ---
  horizontal solid  line ---  along with its  observed error  bars ---
  horizontal  dashed   lines  ---  is  also   displayed.   The  formal
  theoretical errors are also shown as dashed lines.}
\label{fig3}
\end{figure}

\section{Pulsating white dwarfs}

Pulsations   in    white   dwarfs   are   associated    to   nonradial
$g$(gravity)-modes which are a subclass of spheroidal modes whose main
restoring  force is  gravity.  These  modes are  characterized by  low
oscillation frequencies  (long periods) and  by a displacement  of the
stellar fluid  essentially in  the horizontal direction.   Hence, some
characteristics of the  pulsations are sensitive to  the precise value
of $G$,  and to its  rate of change. In  particular, it can  be easily
understood  that  measuring  the  rate  of change  of  the  period  is
equivalent to measure the evolutionary  time scale of the white dwarf.
Thus, a slowly varying $G$ should have an impact in the rate of period
change of the observed periods.  There  are two white dwarfs for which
we have  reliable determinations of the  rate of period change  of its
main   periods,  namely   G117$-$B15A  and   R548.  We   performed  an
asteroseismological analysis  of these  white dwarfs  using a  grid of
fully evolutionary DA models \citep{2012MNRAS.420.1462R} characterized
by consistent  chemical profiles for  both the core and  the envelope,
and  covering a  wide  range  of stellar  masses,  thicknesses of  the
hydrogen envelope and effective temperatures.

The pulsation periods for the modes with $\ell  = 1$ and $k = 1, 2, 3$
and $4$ of the asteroseismological model of G117$-$B15A for increasing
values   of  $|\dot{G}/G|$   are   shown  in   the   upper  panel   of
Fig.~\ref{fig2}.  The variation of the periods is negligible, implying
that a  varying $G$  has negligible  effects on  the structure  of the
asteroseismological  model,  and  that,  for  a  fixed  value  of  the
effective temperature,  the pulsation periods are  largely independent
of  the  adopted  value  of  $|\dot{G}/G|$.  In  the  lower  panel  of
Fig.~\ref{fig2} we  display the  rates of period  change for  the same
modes.   At odds  with what  happens with  the pulsation  periods, the
rates  of  period change  are  markedly  affected  by a  varying  $G$,
substantially increasing for increasing values of $|\dot{G}/G|$.  This
is because  that, for a decreasing  value of $G$ with  time, the white
dwarf  cooling  process  accelerates \citep{gold,gnew},  and  this  is
translated into  a larger secular  change of the pulsation  periods as
compared with the situation in which $G$ is constant.

In Fig.~\ref{fig3}  we plot the  theoretical value of  $\dot{\Pi}$ the
mode with period $\Pi = 215$~s of G117$-$B15A for increasing values of
$|\dot{G}/G|$ (solid  curve).  The  dashed curves embracing  the solid
curve show  the uncertainty in  the theoretical value  of $\dot{\Pi}$,
$\epsilon_{\dot{\Pi}} =  0.09 \times 10^{-15}$ s  s$^{-1}$. This value
has been derived  taking into account the uncertainty due  to our lack
of knowledge of the  $^{12}$C$(\alpha,\gamma)^{16}$O reaction rate ---
$\varepsilon_1 \sim 0.03 \times 10^{-15}$~s/s  --- and that due to the
errors in  the asteroseismological model ---  $\varepsilon_2 \sim 0.06
\times 10^{-15}$~s/s \cite{2012MNRAS.424.2792C}.   We assumed that the
uncertainty for the case in which $\dot{G} \neq 0$ is the same as that
computed  for  the  case  in  which $G=  0$,  which  is  a  reasonable
assumption.   Considering that  the  theoretical  solution should  not
deviate more than one standard deviation from the observational value,
we conclude  that the secular  rate of variation of  the gravitational
constant obtained  using the  variable DA  white dwarf  G117$-$B15A is
$\dot{G}/G=    (-1.79^{+0.53}_{-0.49})   \times    10^{-10}$~yr$^{-1}$
\citep{varG}.   The same  analysis applied  to the  other star,  R548,
results     in      $\dot{G}/G=     (-1.29^{+0.71}_{-0.63})     \times
10^{-10}$~yr$^{-1}$, a  very similar  upper bound ---  see \cite{varG}
for  a  detailed discussion.   Clearly,  these  values are  completely
compatible each other, although  currently less restrictive than those
obtained using other techniques, and compatible with a null result for
$\dot G/G$.

\section{Discussion, conclusions and outlook}

In  this work  we have  reviewed  the several  constraints that  white
dwarfs  can provide  on  the rate  of change  of  a secularly  varying
gravitational  constant.  Specifically,  we  have shown  that  when  a
secularly evolving  value of $G$  is adopted, the white  dwarf cooling
tracks (and the main sequence  evolutionary times of their progenitors
as well) are  noticeably affected, and depend sensitively  not only on
the value of $\dot  G/G$ but also on the actual value  of $G$ when the
white dwarf  was born. We find  that for negative  values of $\dot  G$ the
cooling is accelerated, due  a less intense gravitational interaction.
According to these results the main  sequence turn-off age and the age
derived  from the  termination  of the  white  dwarf cooling  sequence
differ from those computed when  a constant value of Newton's constant
is adopted. This can be used to  constrain the rate of variation of a
rolling  $G$. In  particular, we  have applied  this technique  to the
metal rich,  well populated, old  open cluster NGC~6791. It  turns out
that the resulting age of  NGC~6791 is considerably modified modified,
as it  occurs with the position  of the main sequence  turn-off in the
color-magnitude  diagram,  if the  same  distance  modulus is  adopted.
Accordingly, the distance modulus necessary to fit the position in the
color-magnitude diagram of  the main sequence turn-off  of the cluster
needs to  be changed as  well.  However, the distance  modulus derived
using an  independent and  reliable method (eclipsing  binaries) which
does not make use of theoretical models turns out to be $13.46\pm 0.1$
\citep{Grundahl}.  Thus, large errors in  the distance modulus seem to
be quite implausible.   Given the uncertainty in  the distance modulus
($\simeq 0.1^{\rm  mag}$), and the  measured value, an upper  limit to
$\dot  G/G$   can  be  placed.   Since   $\Delta  t_{\rm  MSTO}/\Delta
(m-M)_{\rm F606W}\approx  4$~Gyr/mag, the maximum age  difference with
respect to the case in which a constant $G$ is adopted should be $\sim
0.4$~Gyr,  which   translates  into  an  upper   bound  $\dot  G/G\sim
-1.8\times   10^{-12}$~yr$^{-1}$.   This   upper  limit   considerably
improves the other  existing upper bounds to the rate  of variation of
$G$ and is equivalent to the upper limit set by helioseismology.

We  have  also shown  that  individual  pulsating hydrogen-rich  white
dwarfs  can also  be useful  in setting  upper limits  to the  rate of
variation  of $G$,  although  these upper  bounds  are currently  less
restrictive than  those obtained using the  color-magnitude diagram of
clusters. In essence,  we have found that the periods  of the dominant
modes of the two white dwarfs  (G117--B15A and R548) for which we have
reliable observational  determinations of their rate  of period change
of their  main modes,  are not  affected, thus  allowing to  derive an
excellent    asteroseismological    fit     of    their    pulsational
spectra. However, the  rates of period change of  their dominant modes
are severely affected, as a consequence  of the fact that the rates of
change of these  periods reflect their evolutionary  changes, and thus
allow to measure  their evolutionary time scales.  Accordingly, in the
case  of  an  hypothetical   smooth  variation  of  the  gravitational
constant, and due to the sensitivity of the cooling time scales to the
precise  value   of  $\dot  G$,   the  rates  of  period   change  are
modified. Our calculations in the case of a running value of $G$ allow
to  compare  the  predictions  of  the  theoretical  models  with  the
observational rates of period change,  and hence to derive constraints
on the  value of $\dot G/G$.   Using this technique we  found that the
upper   bounds   are    $\dot{G}/G=   (-1.79^{+0.53}_{-0.49})   \times
10^{-10}$~yr$^{-1}$,  and  $\dot{G}/G= (-1.29^{+0.71}_{-0.63})  \times
10^{-10}$~yr$^{-1}$,   for  G117--B15A   and  R548   respectively.  We
emphasize that although  these upper limits are  less restrictive than
those obtained using the previously described technique, they could be
much improved should we  have reliable observational determinations of
the rate of change of the  dominant periods of pulsating massive white
dwarfs, as the effect of a running $G$ is more evident for these white
dwarfs, due to their larger gravitational field.

Last, but  not least, we  would like to  emphasize here that  there is
still room for new (and possibly  exciting) studies that have not been
addressed here, and  that the results of such  studies could translate
in  interesting improved  constraints.   To be  precise,  we now  have
excellent  observational  luminosity  functions  of  the  white  dwarf
population  of  the  Galactic  disk,  which are  the  result  of  both
magnitude-limited large scale  surveys --- like the  Sloan Digital Sky
Survey   \citep{SDSS1,SDSS2}    or   the   SuperCOSMOS    sky   survey
\citep{SuperCOSMOS}     ---    or     of    volume-limited     surveys
\citep{Bergeron}. The completenesses of  the large surveys is expected
to be high  ($\sim 80\%$), while the volume-limited  sample is thought
to be  nearly complete. The  white dwarf luminosity  function reflects
the characteristic cooling time of the population of white dwarfs as a
function of the absolute bolometric magnitude, and has two distinctive
features.  The first of these properties is a monotonic increase until
luminosities of  the order of  $\log (L/L_\odot) \sim -3.5$,  which is
simply a consequence of the the fact that due to the absence of energy
sources other than the gravothermal energy of white dwarfs, the cooler
a white dwarf the longer it takes  to cool further. The second --- and
for our  purposes most important feature  --- of the disk  white dwarf
luminosity   function   is   the   presence  a   sharp   drop-off   at
$\log(L/_\odot) \sim  -4.5$. This  pronounced cut-off is  the obvious
consequence of the finite age of the Galactic disk. The origin of this
deficit of cool stars is clear:  white dwarfs have not had time enough
to cool  down beyond  this luminosity.  Since  the cooling  process of
white dwarfs is  sensitive to $\dot G/G$ it is  rather evident that in
the case of a secularly varying $G$ the position of the cut-off should
be  different. Such  an  analysis  still remains  to  be  done, as  it
requires the calculation of an  extensive set of cooling sequences for
different values  of $\dot G/G$  and the  initial value of  $G$, which
requires considerable efforts, but it is one of our priorities for the
next future.

In summary, we have demonstrated that due to their relative structural
simplicity, to  the fact  that the  gravothermal cooling  process that
governs their  evolution is  well understood,  to the  well determined
individual and  ensemble properties, and  to the sensitivity  of their
properties to the  value of $\dot G/G$, white dwarf  stars can be used
to  constrain alternative  theories  of gravitation,  and that  future
efforts, both on  the observational and on the  theoretical sides, can
result  in  improved  upper  bounds  on the  rate  of  change  of  the
gravitational constant.

\begin{acknowledgements}
Part of  this work was  supported by  AGENCIA through the  Programa de
Modernizaci\'on Tecnol\'ogica BID  1728/OC-AR, by PIP 112-200801-00940
grant from CONICET, by MCINN grant AYA2011-23102, by the ESF EUROCORES
Program  EuroGENESIS (MICINN  grant  EUI2009???04170),  by the  European
Union FEDER funds, and by the AGAUR.
\end{acknowledgements}

\bibliographystyle{aa}
\bibliography{Sext}

\begin{thebibliography}{25}
\expandafter\ifx\csname natexlab\endcsname\relax\def\natexlab#1{#1}\fi

\bibitem[{{Althaus} {et~al.}(2011){Althaus}, {C{\'o}rsico}, {Torres},
  {Lor{\'e}n-Aguilar}, {Isern}, \& {Garc{\'{\i}}a-Berro}}]{gnew}
{Althaus}, L.~G., {C{\'o}rsico}, A.~H., {Torres}, S., {et~al.} 2011, \aap, 527,
  A72

\bibitem[{{Althaus} {et~al.}(2010){Althaus}, {Garc{\'{\i}}a-Berro}, {Renedo},
  {Isern}, {C{\'o}rsico}, \& {Rohrmann}}]{Althaus10}
{Althaus}, L.~G., {Garc{\'{\i}}a-Berro}, E., {Renedo}, I., {et~al.} 2010, \apj,
  719, 612

\bibitem[{{Bambi} {et~al.}(2005){Bambi}, {Giannotti}, \& {Villante}}]{B05}
{Bambi}, C., {Giannotti}, M., \& {Villante}, F.~L. 2005, \prd, 71, 123524

\bibitem[{{Copi} {et~al.}(2004){Copi}, {Davis}, \& {Krauss}}]{CO4}
{Copi}, C.~J., {Davis}, A.~N., \& {Krauss}, L.~M. 2004, \prl, 92, 171301

\bibitem[{{C{\'o}rsico} {et~al.}(2013){C{\'o}rsico}, {Althaus},
  {Garc{\'{\i}}a-Berro}, \& {Romero}}]{varG}
{C{\'o}rsico}, A.~H., {Althaus}, L.~G., {Garc{\'{\i}}a-Berro}, E., \& {Romero},
  A.~D. 2013, J. Cosm. Astropart. Phys., 6, 32

\bibitem[{{C{\'o}rsico} {et~al.}(2012){C{\'o}rsico}, {Althaus}, {Miller
  Bertolami}, {Romero}, {Garc{\'{\i}}a-Berro}, {Isern}, \&
  {Kepler}}]{2012MNRAS.424.2792C}
{C{\'o}rsico}, A.~H., {Althaus}, L.~G., {Miller Bertolami}, M.~M., {et~al.}
  2012, \mnras, 424, 2792

\bibitem[{{De Gennaro} {et~al.}(2008){De Gennaro}, {von Hippel}, {Winget},
  {Kepler}, {Nitta}, {Koester}, \& {Althaus}}]{SDSS1}
{De Gennaro}, S., {von Hippel}, T., {Winget}, D.~E., {et~al.} 2008, \aj, 135, 1

\bibitem[{{degl'Innocenti} {et~al.}(1996){degl'Innocenti}, {Fiorentini},
  {Raffelt}, {Ricci}, \& {Weiss}}]{Weiss}
{degl'Innocenti}, S., {Fiorentini}, G., {Raffelt}, G.~G., {Ricci}, B., \&
  {Weiss}, A. 1996, \aap, 312, 345

\bibitem[{{Garc{\'{\i}}a-Berro} {et~al.}(2008){Garc{\'{\i}}a-Berro}, {Althaus},
  {C{\'o}rsico}, \& {Isern}}]{GB08}
{Garc{\'{\i}}a-Berro}, E., {Althaus}, L.~G., {C{\'o}rsico}, A.~H., \& {Isern},
  J. 2008, \apj, 677, 473

\bibitem[{{Garc{\'{\i}}a-Berro} {et~al.}(1995){Garc{\'{\i}}a-Berro}, {Hernanz},
  {Isern}, \& {Mochkovitch}}]{gold}
{Garc{\'{\i}}a-Berro}, E., {Hernanz}, M., {Isern}, J., \& {Mochkovitch}, R.
  1995, \mnras, 277, 801

\bibitem[{{Garc{\'{\i}}a-Berro} {et~al.}(2007){Garc{\'{\i}}a-Berro}, {Isern},
  \& {Kubyshin}}]{mio}
{Garc{\'{\i}}a-Berro}, E., {Isern}, J., \& {Kubyshin}, Y.~A. 2007, \aapr, 14,
  113

\bibitem[{{Garc{\'{\i}}a-Berro} {et~al.}(2006){Garc{\'{\i}}a-Berro},
  {Kubyshin}, {Lor{\'e}n-Aguilar}, \& {Isern}}]{IJMPD}
{Garc{\'{\i}}a-Berro}, E., {Kubyshin}, Y., {Lor{\'e}n-Aguilar}, P., \& {Isern},
  J. 2006, Int. J. Mod. Phys. D, 15, 1163

\bibitem[{{Garc{\'{\i}}a-Berro} {et~al.}(2011){Garc{\'{\i}}a-Berro},
  {Lor{\'e}n-Aguilar}, {Torres}, {Althaus}, \& {Isern}}]{JCAP}
{Garc{\'{\i}}a-Berro}, E., {Lor{\'e}n-Aguilar}, P., {Torres}, S., {Althaus},
  L.~G., \& {Isern}, J. 2011, J. Cosm. Astropart. Phys., 5, 21

\bibitem[{{Garc{\'{\i}}a-Berro} {et~al.}(2010){Garc{\'{\i}}a-Berro}, {Torres},
  {Althaus}, {Renedo}, {Lor{\'e}n-Aguilar}, {C{\'o}rsico}, {Rohrmann},
  {Salaris}, \& {Isern}}]{nature}
{Garc{\'{\i}}a-Berro}, E., {Torres}, S., {Althaus}, L.~G., {et~al.} 2010, \nat,
  465, 194

\bibitem[{{Gazta{\~n}aga} {et~al.}(2002){Gazta{\~n}aga}, {Garc{\'{\i}}a-Berro},
  {Isern}, {Bravo}, \& {Dom{\'{\i}}nguez}}]{SNIa}
{Gazta{\~n}aga}, E., {Garc{\'{\i}}a-Berro}, E., {Isern}, J., {Bravo}, E., \&
  {Dom{\'{\i}}nguez}, I. 2002, \prd, 65, 023506

\bibitem[{{Giammichele} {et~al.}(2012){Giammichele}, {Bergeron}, \&
  {Dufour}}]{Bergeron}
{Giammichele}, N., {Bergeron}, P., \& {Dufour}, P. 2012, \apjs, 199, 29

\bibitem[{{Grundahl} {et~al.}(2008){Grundahl}, {Clausen}, {Hardis}, \&
  {Frandsen}}]{Grundahl}
{Grundahl}, F., {Clausen}, J.~V., {Hardis}, S., \& {Frandsen}, S. 2008, \aap,
  492, 171

\bibitem[{{Guenther} {et~al.}(1998){Guenther}, {Krauss}, \&
  {Demarque}}]{Demarque}
{Guenther}, D.~B., {Krauss}, L.~M., \& {Demarque}, P. 1998, \apj, 498, 871

\bibitem[{{Harris} {et~al.}(2006){Harris}, {Munn}, {Kilic}, {Liebert},
  {Williams}, {von Hippel}, {Levine}, {Monet}, {Eisenstein}, {Kleinman},
  {Metcalfe}, {Nitta}, {Winget}, {Brinkmann}, {Fukugita}, {Knapp}, {Lupton},
  {Smith}, \& {Schneider}}]{SDSS2}
{Harris}, H.~C., {Munn}, J.~A., {Kilic}, M., {et~al.} 2006, \aj, 131, 571

\bibitem[{{Hofmann} {et~al.}(2010){Hofmann}, {M{\"u}ller}, \& {Biskupek}}]{H10}
{Hofmann}, F., {M{\"u}ller}, J., \& {Biskupek}, L. 2010, \aap, 522, L5

\bibitem[{{Lor{\'e}n-Aguilar} {et~al.}(2003){Lor{\'e}n-Aguilar},
  {Garc{\'{\i}}a-Berro}, {Isern}, \& {Kubyshin}}]{LAea}
{Lor{\'e}n-Aguilar}, P., {Garc{\'{\i}}a-Berro}, E., {Isern}, J., \& {Kubyshin},
  Y.~A. 2003, Class. \& Quantum Grav., 20, 3885

\bibitem[{{Renedo} {et~al.}(2010){Renedo}, {Althaus}, {Miller Bertolami},
  {Romero}, {C{\'o}rsico}, {Rohrmann}, \& {Garc{\'{\i}}a-Berro}}]{Renedo10}
{Renedo}, I., {Althaus}, L.~G., {Miller Bertolami}, M.~M., {et~al.} 2010, \apj,
  717, 183

\bibitem[{{Romero} {et~al.}(2012){Romero}, {C{\'o}rsico}, {Althaus}, {Kepler},
  {Castanheira}, \& {Miller Bertolami}}]{2012MNRAS.420.1462R}
{Romero}, A.~D., {C{\'o}rsico}, A.~H., {Althaus}, L.~G., {et~al.} 2012, \mnras,
  420, 1462

\bibitem[{{Rowell} \& {Hambly}(2011)}]{SuperCOSMOS}
{Rowell}, N. \& {Hambly}, N.~C. 2011, \mnras, 417, 93

\bibitem[{{Uzan}(2003)}]{uzan}
{Uzan}, J.-P. 2003, Rev. Mod. Phys., 75, 403

\end{thebibliography}

\end{document}